\documentstyle[11pt,pasp,twoside]{article}
\def\edcomment#1{\iffalse\marginpar{\raggedright\sl#1\/}\else\relax\fi}
\reversemarginpar
\begin{document}
\title{Observations of Galaxies with Future X-ray Observatories}
\author{Giuseppina Fabbiano} 
\affil{Harvard-Smithsonian Center for Astrophysics, 60 Garden
St., Cambridge MA 02138 USA}

%%%%%%%%%%%%%%%%
\begin{abstract}
Normal galaxies are faint and complex X-ray sources that provide
very powerful probes for fundamental astrophysical questions. Examples
include: the study of populations of X-ray emitting sources;
the study of the entire spectrum of black-hole phenomena; 
and galaxy formation and evolution in interaction with the surrounding environment.
While exciting, all these fields {\bf require} Chandra sub-arcsecond
resolution, and at least comparable spectral capabilities. They also
require much large collecting areas. I argue that this is the
direction we must plan for future X-ray observatories.

\end{abstract}
%%%%%%%%%%%%%%%%

%%%%%%%%%%%%%%%%%%%%%%%%%%%%%%%%%%%%%%%%%%%%%%%%%%%%%%%%%%%%%%%%
\section{Introduction: the X-ray emission of galaxies}

Normal galaxies (i.e. those not dominated by an AGN) are relatively faint 
X-ray sources, in most cases fainter than the detection threshold attainable with 
non-imaging X-ray observatories. It is therefore not surprising that
the study of their X-ray properties became possible only in the late 1970s-early
1980s, after the launch of the {\it Einstein Observatory}, the first 
imaging X-ray telescope (Giacconi et al 1979).

Strange as it seems now, at the time there was a fairly widespread misconception that
galaxies would be `boring' targets of X-ray observations, giving us
information only on the collective emission of Milky-Way-like populations
of accreting binaries and SNRs. Fortunately, the data soon proved this wrong.

We now know that X-ray sources in all types of galaxies include not only XRB and SNR
(and these are very interesting on their own right as probes into
extreme forms of matter), but also a diffuse hot gaseous interstellar 
medium, either gravitationally bound as in X-ray luminous E galaxies,
or in partial outflow or winds, as for example in the starburst galaxy M82. 
Moreover, observations of galaxies are essential for studying the whole gamut 
of nuclear emission, extending from QSOs to low-luminosity nuclei. This includes
starburst nuclear regions, LINERS and low-luminosity AGN (see reviews, Fabbiano 1989, 1996).
With X-ray observations we can probe the physics of these
components and their astrophysical implications for 
galaxy formation and evolution as well as cosmology.

All this knowledge was the result of the {\it Einstein} observations,
followed by the higher angular resolution (but softer energy band) ROSAT,
and by the wide-energy-band (but larger-beam) satellites ASCA and BeppoSAX.

Now, with Chandra and XMM-Newton, this field is being pushed to new
heights. In what follows, I will summarize the principal scientific themes
for which X-ray studies of galaxies are relevant, and I will then 
point out the implied requirements for future X-ray observatories.
Previous papers discussing this type of requirements, which are still
valid, are Fabbiano (1990) and Elvis \& Fabbiano (1997).

%%%%%%%%%%%%%%%%%%%%%%%%%%%%%%%%%%%%%%%%%%%%%%%%%%%%%%%%%%%%%%%%
\section{Science Themes}

Galaxies can be considered test particles tracing the large-scale distribution of
matter in the Universe, but they are also complex `living' organisms. 
They form, they age, they interact and merge, they rejuvenate, they
interact with and modify their environment. X-ray observations 
can provide precious and unique insight into this life-cycle, and into
the life-cycle of the galaxian components.

\subsection{Populations of X-ray sources: formation and evolution}

It is well known that strong X-ray emission may arise from stars
in the final stages of their evolution, either from Supernova Remnants (SNR)
or from binary systems (XRB) including a white dwarf, a neutron star, or a black 
hole. Galactic XRBs and SNRs were among the first types of X-ray source 
discovered in X-rays and have been studied intensely since the UHURU days
(see Giacconi \& Gursky 1974). 
The study of Galactic X-ray sources has provided a wealth of
information on these systems and a good understanding of the nature of individual
objects, but questions about their population properties, relations to their
galaxian environment
and stellar populations, and evolution can be answered only by observing a 
wide range of external galaxies.

While this type of work had been attempted with {\it Einstein} and ROSAT
for M31 and a few other nearby galaxies (see Fabbiano 1995), it is only with Chandra
that this field is starting to blossom. The reason is simple: angular resolution.
Chandra's sub-arcsecond resolution corresponds to physical sizes of $\sim 70$~pc
at a distance of $\sim30$~Mpc (and $\sim 35$~pc at the Virgo Cluster). Given
the relative sparse distribution of luminous X-ray sources, in most cases this
resolution  allows the detection of individual bright X-ray sources at 
least out to these distances. Moreover, because of the very small Chandra beam,
only a few photons are needed for a significant detection.  As a result,
X-ray luminosity functions (XLF) of galaxian sources have recently been
derived with unprecedented sensitivity from Chandra observations. 
Early results include the XLF of NGC~4697 (Sarazin, Irwin \& Bregman 2000) 
an X-ray-faint elliptical 
galaxy at a distance of 16~Mpc, where for the first time the XRB
population was directly detected. In this XLF, a break near $3 \times
10^{38}$~ergs/s was reported and interpreted as the signature of a population
of black-hole binaries. In M81, at a distance of 3.5~Mpc,  the XLF can be followed
down to luminosities of $\sim 3 \times 10^{37}$ergs/s, 
and different distributions are found for bulge
and disk sources (Tennant et al 2001). The XLFs of the starburst galaxy M82 and of the merging
galaxies NGC~4038/9 (Zezas et al 2001) show significant extensions to very 
high luminosities, well above those seen in Milky Way sources, suggesting
the presence of a short-lived population of very luminous XRBs, possibly 
associated with black holes of masses surpassing those of Galactic black-hole
binaries.  

These results are exciting, because they provide a very poweful new tool for
studying the population of X-ray sources and from this inferring constraints 
either on their formation history as it relates to intense star formation
episodes in the mother galaxy (e.g. Wu 2001), or on the nature of some of these
sources. 

However, we must remember that Chandra is a small telescope, so that prohibitive
observing times will be needed to obtain XLFs from a large enough sample of galaxies,
extending down to luminosities well in the range of normal XRBs. Moreover, while
10 photons make a very significant source, they cannot give us any useful information
on the spectral signature of the emission. The latter is needed to identify different types
of X-ray sources, and in particular to separate SNRs from XRBs.
XMM, with its larger collecting area,
will allow a systematic spectral study of the populations of X-ray sources in
the very nearby universe, but source confusion (comparable or worse than that
of the ROSAT HRI) will be the limiting factor for distances larger than a few 
megaparsecs.

If we want to continue populations studies in the future, and extend them to a
large number of galaxies and to the spectral dimension, we need to consider an
X-ray telescope that retains Chandra's imaging and spectral capabilities, while
providing a significantly larger collecting area.

\subsection{Small to Humongous Black Holes}

Studying Black Holes and the range of phenomena associated with the interaction 
of matter with Black Holes is certainly an exciting field, and a field to which
X-ray observations can give unique contributions. It is generally believed that
luminous AGN are the result of accretion onto massive ($\sim 10^{7-8}~M_{\odot}$)
nuclear black holes, and dynamical evidence now points to the widespread association
of such black holes with galaxian bulges (Magorrian et al 1998). 
At the other end of the spectrum 
of black hole masses, black holes of a few solar masses have been identified as
the compact source in some Galactic XRBs (see Tanaka \& Lewin 1995). 
These two categories of black holes
may have different origins (albeit always involving gravitational collapse):
the massive nuclear black holes may either be primordial, or be the result of
accretion onto a nuclear seed black hole during the life time of the host galaxy;
the few-solar-mass black holes in XRBs are the likely product of the evolution
and final collapse of a star too massive to end up as a neutron star.

But is this the entire spectrum of black hole masses? Are there any intermediate mass
black holes (in the $\sim 100~M_{\odot}$ range)?
This type of object had been hypothesized as the accretor in exceptionally
luminous sources first discovered in external galaxies with {\it Einstein},
and named `Super-Eddington sources', because their X-ray luminosity was well
in excess of that of accretion onto a solar-mass neutron star (see Fabbiano 1995).
ASCA monitoring revealed spectral and spectral/temporal 
characteristics reminiscent of black hole Galactic XRBs (e.g. Makishima et al 2000;
Kubota et al 2001) in a few of these sources, but for most of them, even with 
{\it Einstein}'s and ROSAT's 5'' resolution, it could not be excluded that
the emission arose from extended regions or clumps of more normal sources.

With Chandra, the census of `Super-Eddington sources' is increasing dramatically. 
These sources have been separated out and 
detected in large numbers in previously confused bright
emission regions associated with intense star formation activity (e.g. in M82,
Matsumoto et al 2001, Kaaret et al 2001;
the Antennae galaxies, Fabbiano, Zezas \& Murray 2001; NGC~253, Weaver, Strickland \&
Heckman 2001), and what previously appeared as an 
interesting but relatively rare occurrence is now a widespread phenomenon.
Moreover, the luminosities of some of these sources would imply black holes 
in the $\geq 100~M_{\odot}$ range if spherical accretion applies. 
Are these sources really this massive? And if so, how did they form? A single stellar
progenitor  may require an unrealistically high mass. Could these objects result from
direct collapse of massive molecular clouds, or from accretion onto seed black holes 
in compact star clusters? Or perhaps these sources are not that massive after all
and beaming in a fraction of `normal' XRBs could explain their very high 
luminosities (King et al 2001).
Clearly this is an exciting field, that requires follow-up and in-depth study,
both from the point of view of deriving `deep' XLFs of galaxies to model
different X-ray source populations, and for constraining spectral and time
variability signatures.

Other examples, also related to the physics of accretion on black holes, 
which are now opening up with Chandra, include: (1) the
search for the X-ray-faint conterparts of quiescent nuclear massive black holes
(e.g. Di Matteo, Carilli \&  Fabian 2001);
(2) the study of  circumnuclear regions, and of the interplay between starburst and AGN
phenomena (Ogle et al 2000).

These are all examples of requirements for a large area telescope, that
otherwise preserves Chandra's angular resolution.
High angular resolution and sensitivity are needed to detect faint sources
in crowded regions; collecting power is needed 
so that spectral and timing data can be obtained and used for constraining emission models.

\subsection{Galaxy Formation and Evolution}

X-ray observations give us a unique way to explore phenomena 
occurring during galaxy formation, that are important for the evolution 
of both galaxies and their surrounding medium.

Besides the populations of luminous X-ray sources
discussed in the previous sections, 
star formation activity in nearby galaxies 
produces a hot ISM, that in most extreme cases escapes
from the parent galaxy with galactic-scale superwinds (e.g Fabbiano 1988; 
Lehnert \& Heckman 1996).
It is not difficult to imagine that if these winds occur in relatively
minor local examples of this phenomenon, they must have been
much more powerful and widespread at the epoch of the collapse of
the primordial pre-galaxian cloud and the ensuing violent formation 
of early generations of stars. Vestigial evidence of these
winds exists in the presence of metals in the intracluster medium,
that was revealed by X-ray spectroscopy (e.g. Fukazawa et al 2000).
Superwinds are also now being detected with the Keck telescope in the
optical spectra of high redshift galaxies (Shapley et al 2000).
In addition to star formation, nuclear activity may also 
trigger superwinds in elliptical galaxies (Ciotti \& Ostriker 2001),
and galaxy interaction and merging may also be a factor (D'Ercole, Recchi \& Ciotti
2000).

Thanks to the results of X-ray observations,
the old concept of `closed-box' models for galaxy 
evolution, where the evolution could be traced only in terms of
stellar evolution in a defined volume of space, is not longer
viable.
More complex `feedback' models and hydrodynamical simulations 
connecting galaxy and cluster (ambient) evolution are being developed (Cavaliere,
Giacconi \& Menci 2000). 
Since the hot superwinds are the main vector by which energy - and
metals - are transferred from galaxies to the surrounding medium,
direct X-ray observations provide unique constraints to these models.
In particular, detailed spatial/spectral X-ray observations of
relatively nearby galaxies can be used as a way to calibrate these
simulations. By themselves and in comparison with similarly resolved
multiwavelength data, X-ray observations are needed for understanding 
the detailed and complex astrophysical phenomena
affecting all the phases of the ISM, and its interaction with 
stellar evolution and magnetic fields (e.g. McKee \& Ostriker 1977; 
Cox \& Anderson 1982).

A detailed study requires spatial and energy resolution, and
collecting power, such that
individual features (clouds, superbubbles) of the hot ISM can be
mapped both spatially and spectrally. Given that Chandra images
of nearby systems reveal typical scales of hundreds of pc for
some of these features (Fabbiano, Zezas \& Murray 2001), a lesser angular resolution would
not allow this type of comparisons. As already remarked, however,
the number of photons required for a spatial/spectral mapping is
above what a reasonably long Chandra observation can provide in
most cases, requiring a significantly larger collecting area.

%%%%%%%%%%%%%%%%%%%%%%%%%%%%%%%%%%%%%%%%%%%%%%%%%%%%%%%%%%%%%%%%
\section{The Future: building upon Chandra}

In science we should always strive to go forward and to explore
new aspects of the observable universe. Because of their sub-arcsecond
angular resolution, joined with spectral resolution, Chandra
observations are resulting in an impressive leap forward in our
understanding of the X-ray properties of galaxies and their components.
Moreover, this angular resolution makes possible a direct comparison
with optical-UV Hubble data and high resolution radio data, opening up
a truly panchromatic understanding of the processes and of the
nature of the X-ray sources. Future comparison with SIRTF results
will expand this to the IR as well.
With Chandra there is definitely the
feeling that X-ray astronomy has fully come of age as a 
way to explore the astrophysical properties of the universe,
rather than being an exploratory discipline for a small congress of
devotees.

How can we {\bf improve} on this?
{\it We must preserve sub-arcsecond angular resolution}, while retaining or
improving the spectral resolution and substantially improving
the telescope collecting area. 

\vspace{2mm}
\noindent
This work was partially supported under NASA contract NAS8-39073 (CXC).

\vspace{-4mm}
%%%%%%%%%%%%%%%%%%%%%%%%%%%%%%%%%%%%%%%%%%%%%%%%

%%%%%%%%%%%%%%%%%%%%%%%%%%%%%%%%%%%%%%%%%%%%%%%%

%%%%%%%%%%%%%%%%

\begin{references}

\reference
Cavaliere, A.; Giacconi, R.; Menci, N. 2000, ApJ(Letters), 528, L77

\reference
Ciotti, L. \&  Ostriker, J. P. 2001, ApJ, 551, 131

\reference
Cox, D. P. \& Anderson, P. R. 1982, ApJ, 253, 268
\reference
McKee, C. F. \&  Ostriker, J. P. 1977, ApJ, 218, 148

\reference
D'Ercole, A., Recchi, S. \& Ciotti, L. 2000, ApJ, 533, 799

\reference
Di Matteo, T., Carilli, C. L. \&  Fabian, A. C. 2001, 547, 731

\reference Elvis M., \& Fabbiano G., 1997, in `Next Generation
X-ray Observatories' eds. M.J.L. Turner and M.G. Watson
[Leicester University X-ray Astronomy Group Special Report,
XRA97/02], p. 33. {\tt astro-ph/9711178}

\reference
Fabbiano, G. 1988, ApJ, 330, 672

\reference
Fabbiano, G. 1989,  Ann. Rev. Ast. Ap., 27, 87

\reference
Fabbiano, G. 1990, in `High Energy Astrophysics in the 21st 
Century', ed. P. C. Joss, p. 74 (New York: AIP)

\reference
Fabbiano, G. 1995, in `X-Ray Binaries',
W. H. G. Lewin, J. van Paradijs, E. P. J. van den Heuvel, eds.,
Cambridge Astrophysics Series (Cambridge: University Press), p. 390

\reference
Fabbiano, G. 1996, in `The Physics
of LINERS in View of Recent Observations', M. Eracleous, A. Koratkar,
C. Leitherer and L. Ho, eds., ASP Conf. Ser. 103, 56

\reference
Fabbiano, G., Zezas, A., \& Murray, S. S. 2001, ApJ, in press

\reference
Fukazawa, Y., Makishima, K., Tamura, T., Nakazawa, K., Ezawa, H., Ikebe, Y., Kikuchi, K. \&
 Ohashi, T. 2000, MNRAS, 313, 21

\reference 
Giacconi, R. et al 1979, ApJ 230, 540

\reference
Giacconi, R. \& Gursky, H. (editors), `X-ray Astronomy', 1974
(Dordrecht: Reidel)

\reference
Kaaret, P., Prestwich, A. H., Zezas, A., Murray, S. S.,
 Kim, D.-W., Kilgard, R. E., Schlegel, E. M., Ward, M. 2001, MNRAS, 321, L29

\reference
King, A. R., Davies, M. B., Ward, M. J., Fabbiano, G. \& Elvis, M
2001, ApJ(Letters), 552, L109

\reference
Kubota, A., Mizuno, T., Makishima, K., Fukazawa, Y., Kotoku, J., Ohnishi, T., Tashiro, M.
2001, ApJ(Letters), 547, L119

\reference
Lehnert, M. D. \& Heckman, T. M. 1996, ApJ(Letters), 462, 651

\reference
Magorrian, J. et al 1997, AJ 115, 2285

\reference
Makishima, K. et al 2000, ApJ, 535, 632

\reference
Matsumoto, H., Tsuru, T. G., Koyama, K., Awaki, H.,
 Canizares, C. R., Kawai, N., Matsushita, S.,
 Kawabe, R. 2001, ApJ(Letters), 547, L25

\reference
Ogle, P. M., Marshall, H. L., Lee, J. C., \& Canizares, C. R. 2000, ApJ(Letters), 545, L81

\reference
Sarazin, C. L., Irwin, J. A., \& Bregman, J. N. 2000, ApJ(Letters) 544, L101

\reference
Shapley, A. E., Steidel, C. C., Adelberger, K. L.,
 Pettini, M., Dickinson, M. E. \& Giavalisco, M. 2000, AAS Abstract, 197, 7714

\reference
Tanaka, Y. \& Lewin, W. H. G. 1995, in in `X-Ray Binaries',
W. H. G. Lewin, J. van Paradijs, E. P. J. van den Heuvel, eds.,
Cambridge Astrophysics Series (Cambridge: University Press), p. 126

\reference
Tennant, A. F., Wu, K., Ghosh, K. K., Kolodziejczak, J. J., Swartz, D. A. 2001,
ApJ(Letters), 549, L43

\reference
Weaver, K. A., Strickland, D. K. \&  Heckman, T. M. 2001, AAS Abstract, 198, 906

\reference
Wu, K. 2001, astro-ph/0103157

\reference
Zezas, A., Fabbiano, G., Rots, A., \& Murray, S. S. 2001, in preparation

\end{references}
\end{document}